\documentclass[aps, prb, reprint, groupedaddress,  floatfix, superscriptaddress, showpacs]{revtex4-1} 
\usepackage[colorlinks,allcolors=blue]{hyperref}
\usepackage{graphicx}
\usepackage{dcolumn}
\usepackage{bm}
\usepackage{braket}

\newcommand\fitwidth{1.0\linewidth}

\begin{document}
\title{\textit{Ab initio} evidence for nonthermal characteristics in ultrafast laser melting}

\author{Chao Lian}
\affiliation{Beijing National Laboratory for Condensed Matter
Physics and Institute of Physics, Chinese Academy of Sciences,
Beijing, 100190, P. R. China}

\author{S. B. Zhang}
\email{zhangs9@rpi.edu}
\affiliation{Department of Physics, Applied Physics, and
Astronomy, Rensselaer Polytechnic Institute, Troy, New York 12180,
USA}

\author{Sheng Meng}
\email{smeng@iphy.ac.cn}
\affiliation{Beijing National Laboratory for Condensed Matter
Physics and Institute of Physics, Chinese Academy of Sciences,
Beijing, 100190, P. R. China}
\affiliation{Collaborative Innovation Center of Quantum
Matter, Beijing, 100190, P. R. China}

\date{\today}
\begin{abstract}
	Laser melting of semiconductors has been observed for almost 40 years; surprisingly, it is not well understood where most theoretical simulations show a laser-induced thermal process. \textit{Ab initio} nonadiabatic simulations based on real-time time-dependent density functional theory reveal intrinsic nonthermal melting of silicon, at a temperature far below the thermal melting temperature of 1680~K.
	Both excitation threshold and time evolution of diffraction intensity agree well with experiment. Nonthermal melting is attributed to excitation-induced drastic changes in bonding electron density, and the subsequent decrease in the melting barrier, rather than lattice heating as previously assumed in the two-temperature models. 
\end{abstract}

\maketitle

\section{Introduction}
Materials can exhibit exotic behaviors and dynamics different from the ground state when excited by laser light. Laser excitation generates ultrafast phenomena and unique condensed phases of matter~\cite{Ostrikov2016}. A popular example is ultrafast melting. Melting within a timescale
of less than a picosecond upon photoexcitation has been ubiquitously observed in
a wide range of semiconductors, including Si~\cite{Shank1983,Harb2006,Harb2008}, Ge~\cite{Siders1999,Cavalleri2000,Sokolowski-Tinten2001}, GaAs~\cite{Glezer1995,Sokolowski-Tinten1995,Roland-prb02}, InSb~\cite{Lindenberg2000,Lindenberg2005,Hillyard2007,Rousse2001,Roland-pssb04}, Ge$_2$Sb$_2$Te$_5$~\cite{Li2011,Waldecker2015}, and, most recently, in two-dimensional materials such as TiSe$_2$~\cite{Mohr-Vorobeva2011,Porer2014} and TaS$_2$~\cite{Hellmann2010,Cho2015a}.

\begin{figure}
\includegraphics[width=\fitwidth]{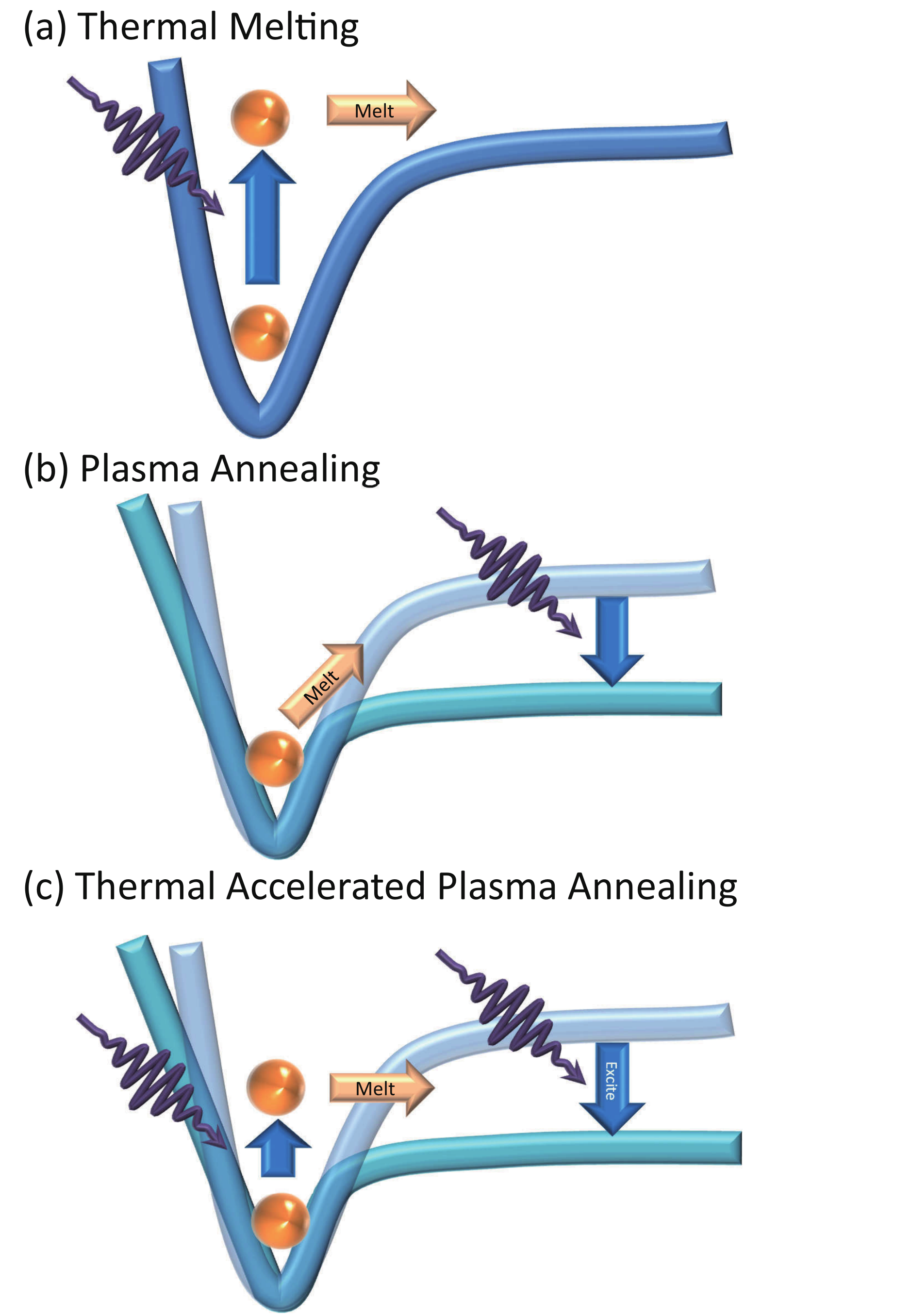}
\caption{\label{NMPAModel} Three models for ultrafast laser
melting: (a) thermal melting, (b) plasma annealing (PA),
and (c) thermal accelerated plasma annealing (TAPA).}
\end{figure}

Despite extensive experimental and theoretical investigations in the past four decades, the atomistic mechanism of ultrafast melting remains controversial, with heavy debates persisting over two representative pictures: thermal melting and plasma annealing (PA)~\cite{VanVechten1979417, VANVECHTEN1979422}, schematically illustrated in Fig.~\ref{NMPAModel}. In experiments plasma annealing, namely, the generation of electron-ion plasma and subsequent bond weakening due to plasma screening, was phenomenologically invoked~\cite{Lindenberg2005,Harb2008}. The PA model assumes that laser energy is completely retained in the electronic subsystem and ultrafast melting is a pure electronic effect. This hypothesis states that the short duration of the melting process is not sufficient for the crystal lattice to be heated and melted~\cite{Sundaram2002}. The model, however, lacks direct evidence from either theory or experiment due to missing information on the lattice temperature and potential energy surface (PES) in excited states. Moreover, the PA model fails to describe the inertial dynamics, namely, melting at constant ion velocities higher than average thermal velocities, as observed experimentally during ultrafast melting \cite{Lindenberg2000}.

First-principles atomistic simulations in principle could provide such information. However, previous theoretical works indicate that a rapid increase in lattice temperature ($10^2$~K $\rightarrow$ $10^3$~K in $100$~fs) is essential to induce melting~\cite{Silvestrelli1996,Ji2013,Medvedev2013,Medvedev2015}, implying that
ultrafast melting is, in fact, a thermal process [Fig.~\ref{NMPAModel}(a)]. However, the state-of-the-art theory and simulation of ultrafast melting suffer from two major limitations, making the result less convincing. The first is the treatment of initial excitation. The two-temperature model (TTM) is prevalent in the literature to describe electronic excitation~\cite{Silvestrelli1996,Zijlstra2008,Zijlstra2013,Ji2013,Hillyard2008,Medvedev2013,Medvedev2015}. The basic assumption is that under laser illumination, the occupation of respective electronic states adopts an equilibrium Fermi-Dirac distribution at the elevated electronic temperature ($T_e\sim10^4$~K), significantly higher than the concurrent ionic temperature of the crystal ($T\sim10^2$~K). However, this assumption conflicts with the fact that hot electrons and holes take $\sim10^2$ to $10^3$~fs to fully relax into a quasiequilibrium state with a well-defined $T_e$~\cite{Sundaram2002}. Thus, $T_e$ is ill-defined and could be irrelevant during ultrafast melting ($\sim100$~fs). Besides introducing the non-Fermi-Dirac distributed electrons in empirical TTM models~\cite{Sun1994,Lisowski2004,Carpene2006}, a new physical model is urgently needed to describe hot electrons and their ultrafast relaxation in a regime far from equilibrium. Another challenge is the inclusion of nonadiabatic effects~\cite{Roland-prb02} such as electron-phonon (el-ph) coupling. Relying on the Born-Oppenheimer (BO) approximation, such effects were often ignored in molecular
dynamics (MD) simulations with TTM~\cite{Silvestrelli1996,Zijlstra2008,Zijlstra2013,Ji2013}, in
spite of a few attempts to introduce empirical parameters to account for nonadiabatic effects~\cite{Medvedev2015,Hillyard2008}. Since nonadiabatic coupling determines the pathway and dynamics of ultrafast energy transfer, a nonadiabatic framework instead of the BO approximation is desirable in simulating ultrafast melting.

In this work, we investigate the atomistic mechanism and dynamics of ultrafast melting of the most popular semiconductor Si, using nonadiabatic molecular dynamics simulations based on real-time time-dependent density functional theory (TDDFT). By solving time-dependent Kohn-Sham equations, TDDFT-MD naturally includes nonadiabatic electron-electron scattering and el-ph effects~\cite{Marques2012}. Optical transitions that reproduce the experimental absorption spectrum are also accounted for. Our parameter-free fully \textit{ab initio} simulations yield intrinsic nonthermal melting dynamics of Si, with an ultrafast time scale ($200$~fs) in a cold lattice ($T<600$~K). The simulated threshold and time evolution of diffraction intensity are almost
identical to experimental measurements. Nonthermal melting is attributed to the decrease in bonding electron density and, in turn, the melting barrier induced by laser illumination, favoring a roughly PA
mechanism. Moreover, we show that small but finite el-ph energy transfer [Fig.~\ref{NMPAModel}(c)], absent in the PA model, is key to inducing accelerated melting dynamics in silicon at low temperatures. This work not only builds a general framework to understand nonthermal melting of semiconductors, but also lays down the foundation for reliable simulations of a wide range of ultrafast physics, now by first principles.

\section{Method}
The calculations are performed with a real-time TDDFT code, time-dependent \textit{ab initio} package (TDAP)~\cite{Meng2008} as implemented in SIESTA~\cite{Ordejon1996,Soler2002,Sanchez-Portal1997}. Crystalline Si is simulated with a supercell of $64$ Si atoms with periodical boundary conditions. The Troullier-Martin pseudopotentials \cite{Troullier1991} and the adiabatic local density approximation \cite{Perdew1981,Yabana1996} for the exchange-correlation functional are used. An auxiliary real-space
grid equivalent to a plane-wave cutoff of $200$~Ry is adopted. The $\Gamma$ point is used to sample the Brillouin zone. During TDDFT-MD the evolution time step is $50$~attosecond for both electrons and ions in a microcanonical ensemble.
\begin{figure}
	\includegraphics[width=\fitwidth]{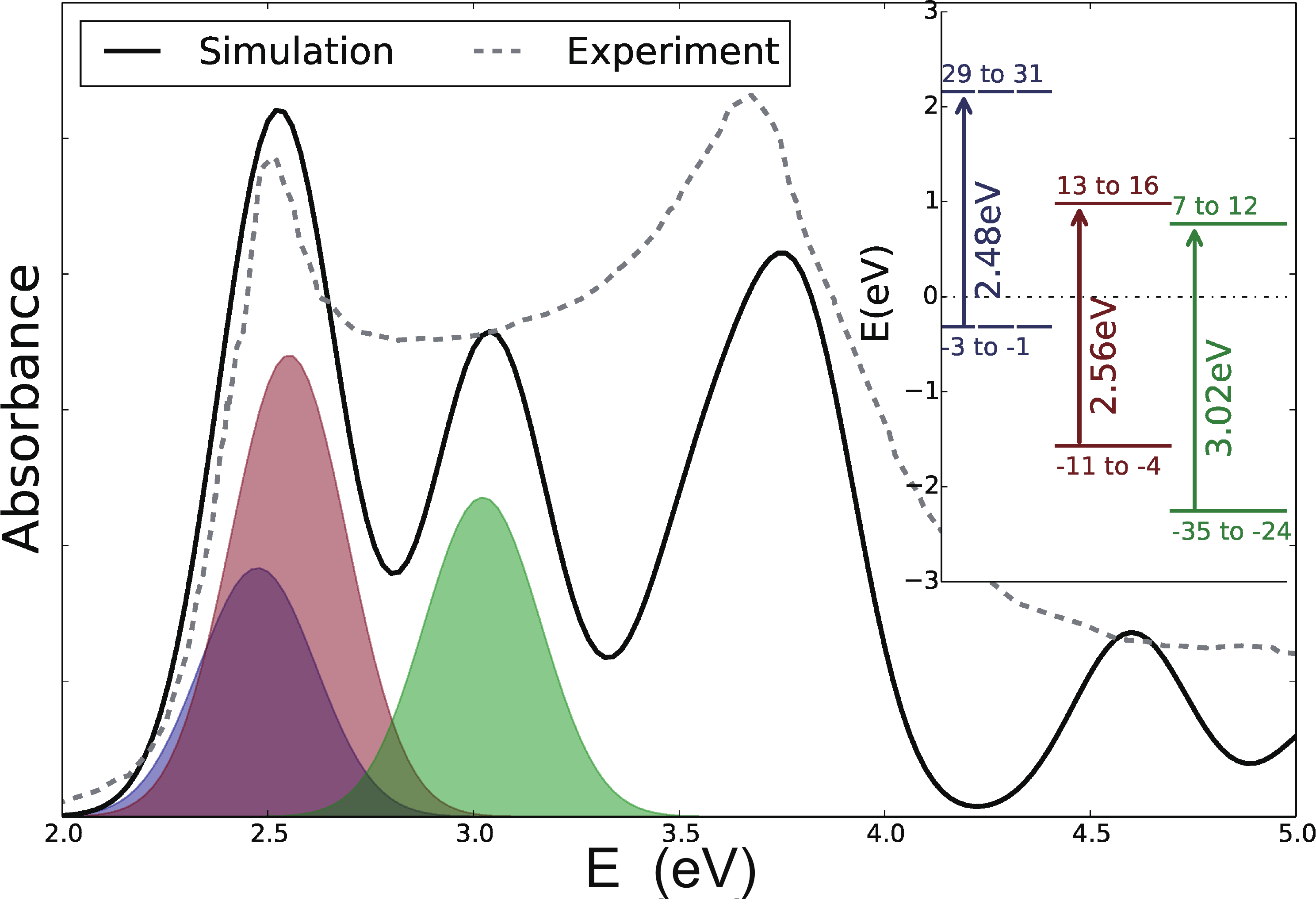}
	\caption{\label{EPSIMG}Optical absorption spectrum (transition
		probability) of Si (solid line). The shadowed zones denote
		transition probabilities of selected excitation modes. The
		dash-dotted line is the experimental spectra red shifted by a scissor
		correction of $0.7$~eV, adopted from Ref. \onlinecite{Lautenschlager1987}.
		Inset: Corresponding electronic bands of selected excitation
		modes. The band indexes are relative to the highest occupied state
		(set as $-1$).}
\end{figure}
In laser-induced phase transitions, the presence of a laser pulse greatly modifies the atomistic dynamics. According to the Runge-Gross theorem \cite{Runge1984}, the initial state is the determinant for many-body dynamics of the electronic system. We thus elaborately build the initial state of photoexcitation by changing the population of Kohn-Sham orbitals from a ground state to
specific configurations. The population variation is proportional to optical transition probability. Laser-induced changes in electron density $\Delta\rho(t_0)$ can be expressed as
\begin{equation}
\label{deltaRho}
\Delta\rho(t_0) \propto \sum_{\left<i,j\right>} P_{ij}
\left(|\psi_i(t_0)|^2 - |\psi_j(t_0)|^2 \right).
\end{equation}
The transition probability $P_{ij}$ from the initial state $\ket{\psi_i}$ to final state $\ket{\psi_j}$ fulfil Fermi's golden rule,
\begin{equation}
P_{ij} = \left|\Braket{\psi_i|\epsilon\cdot
	\hat{P}|\psi_j}\right|^2 \delta\left(\varepsilon_i - \varepsilon_j
- \hbar\omega_{l} \right),
\end{equation}
where $\Braket{\psi_i|\epsilon\cdot \hat{P}|\psi_j}$ is the transition matrix element, $\epsilon$ is the amplitude of the electric field, $\hat{P}$ is the momentum operator, $\omega_{l}$ is the laser
frequency, and $\varepsilon_i$ and $\varepsilon_j$ are the energies of the state $\ket{\psi_i}$ and $\ket{\psi_j}$, respectively. Since DFT usually underestimates the band gap, a scissor correction of
$0.7$~eV is applied when comparing to the experimental absorption spectrum; see Fig.~\ref{EPSIMG}. In experiments, a laser with a wavelength $\lambda_{l} = 387$~nm is used, corresponding to a photon energy of $3.3$~eV. Considering the scissor correction of $0.7$~eV, we use a photon energy of $2.6$~eV in our calculations. The eigenstates involved in these optical transitions are also shown in the inset. Good agreement between the theoretical and experimental absorption spectra is clearly
seen. This method is similar to that proposed by Murray and Fahy
in the study of photoexcitation in Bi \cite{Murray2015}. Schultze \textit{et al}. reported a joint experimental and TDDFT study of the optical absorption process in silicon\cite{Schultze2014a}. In their TDDFT part, an electric field is directly introduced into this system to excite the electrons. Here we use the above simple approach, producing less accurate absorption in silicon, while focusing on ionic dynamics after photoabsorption. The approach nicely reproduces the atomic forces in photoexcited bismuth \cite{Murray2015}. We thus built a physical initial state for photoexcited Si, reflecting consequences after laser absorption. 
  
\section{Results}
\subsection{Laser Melting Under Experimental Conditions} 
\begin{figure}
	\includegraphics[width=0.5\textwidth]{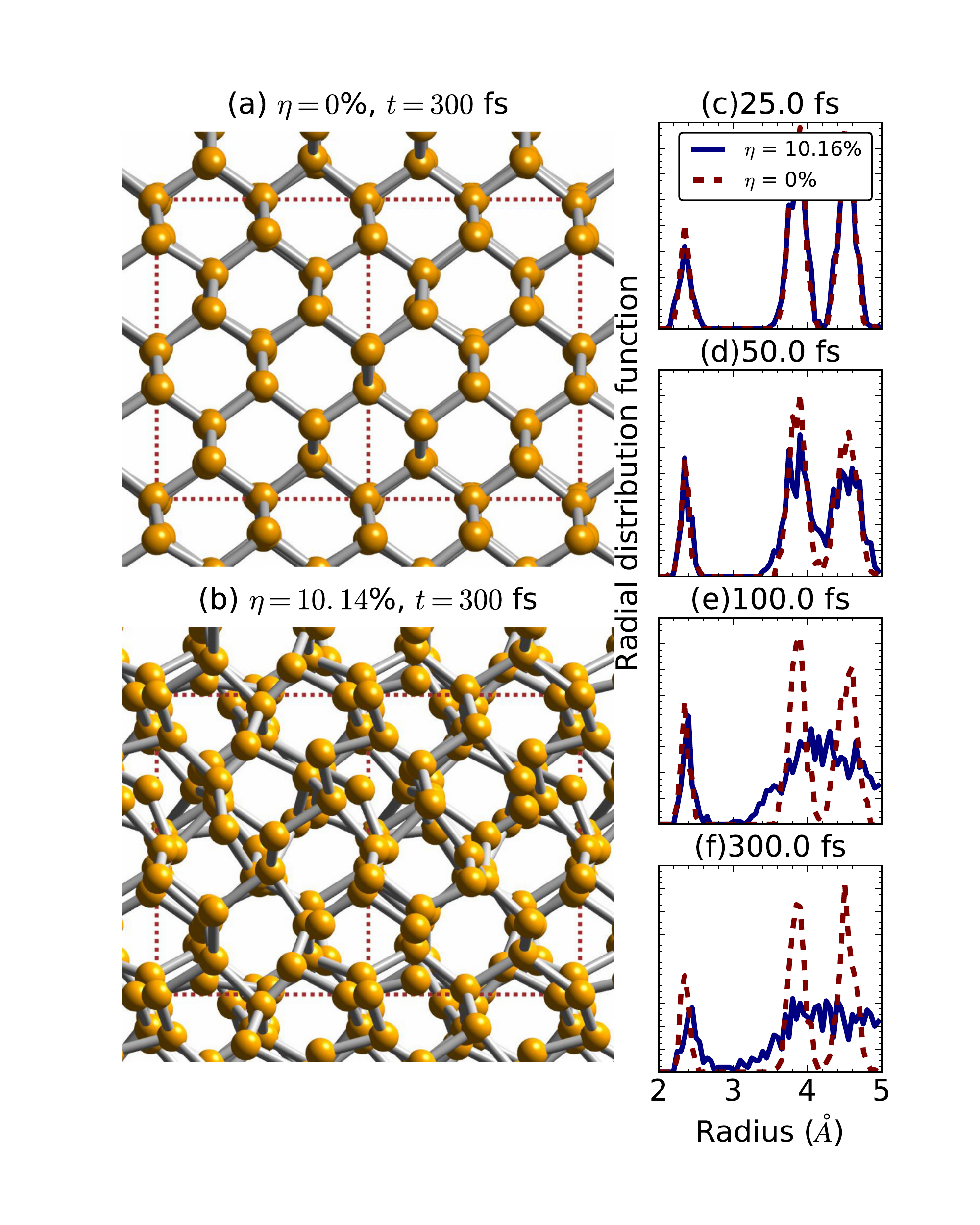}
	\caption{\label{StructRDF}(a-b) The atomic structures during laser melting. (c-e) Evolution of RDF.}
\end{figure}
We simulate laser melting of Si under experimental conditions~\cite{Harb2008}. 
Our simulations based on TDDFT-MD show that without laser illumination, the lattice temperature oscillates around $\sim300$~K because of thermal fluctuations. With laser excitation about $10.16$\% valence electrons are pumped to the conduction bands. Here we use the percentage of valence electrons pumped to denote the laser intensity $\eta$.

As shown in Fig. \ref{StructRDF}, the structures at $t = 300$~fs with $\eta = 10.16$\% and $\eta = 0$\% show the difference caused by the laser. The disorder in Fig. \ref{StructRDF}(b) is clearly caused by the excitation and shows a melting signature. The radial distribution function (RDF) of Si-Si bonds at $t=0$, $25$, $75$ and $275$~fs after excitation with $\eta = 10.16$\% and $\eta = 0$\% shows the same features of melting as in Fig.~\ref{StructRDF}(c)-density increase-\ref{StructRDF}(f): the first peak shifts right, implying the increase in the nearest-neighbor distance; and all peaks become diffusive, indicating a broken crystalline order. 

\begin{figure}
\includegraphics[width=\fitwidth]{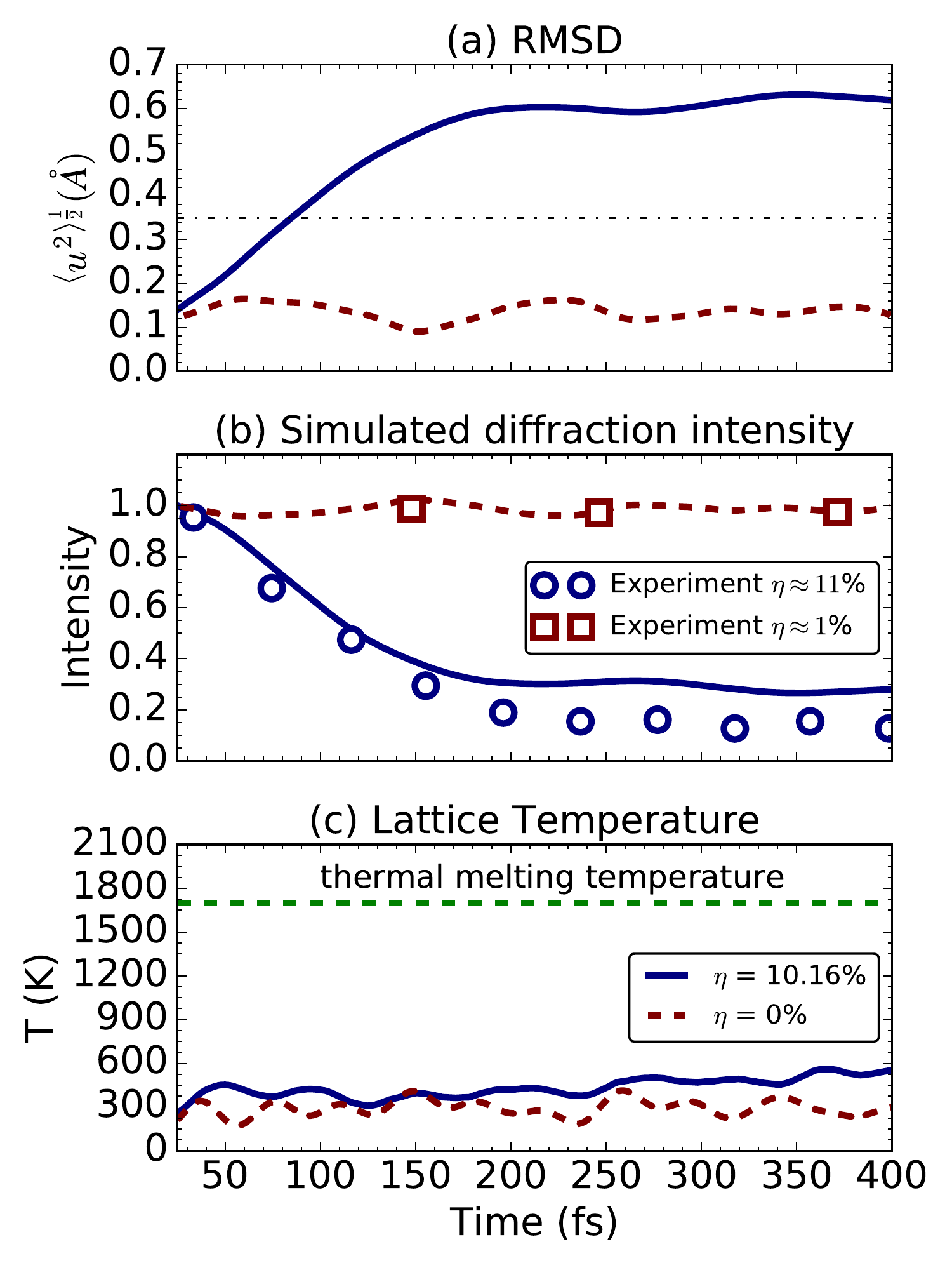}
\caption{\label{T+RMSD+Int}(a) RMSD as a function of
time. (b) Simulated and experimental electron diffraction
intensity of $(220)$ reflection as a function of time.
Experimental data (time rescaled by a factor of 0.33) are taken
from Refs. \onlinecite{Harb2006,Harb2008}. (c) Ionic temperature as a function of time.}
\end{figure}

For a quantitative evaluation, we adopt
the Lindemann criterion: Si melts when its root-mean-square
displacement (RMSD) $\left<u^2(t)\right>^{\frac{1}{2}}$ is larger
than the critical value $R_c = 0.35$~\AA\ \cite{Zijlstra2013}. The
RMSDs with and without laser are shown in Fig.~\ref{T+RMSD+Int}(a). The
maximum RMSD without laser ($\eta = 0$\%) reaches only half of
$R_c$. However, the RMSD with laser intensity $\eta = 10.16$\%
crosses the $R_c$ in $100$~fs and keep increasing to about
$0.6$~\AA\ within $400$~fs, showing an evident ultrafast melting
behavior.

The RMSD is directly connected to diffraction intensity $I(t)$
through the Debye-Waller formula:
\begin{equation}
I(t) = \exp\left[-Q^2\left<u^2(t)\right>/3\right],
\end{equation}
where $Q$ is the reciprocal lattice vector of the probed reflection, $\left<u^2(t)\right>$ is the mean square displacement, i.e. the square of RMSD. We simulate the $I_{\eta=10.16\%}(t)$ and $I_{\eta=0\%}(t)$ for the $(220)$ reflection as shown in Fig.~\ref{T+RMSD+Int}(b). The features of the experimental data \cite{Harb2008} and our simulations are almost identical. The $I_{\eta\sim11\%}(t)$ decreases to $\sim0.2$ after melting in both the experiment and our simulation, while without laser, both the simulated and experimental $I_{\eta\sim0\%}(t)$ shows no drift but an oscillation around 0.95. It demonstrates that our simulation captures the most important features of nonthermal melting observed in experiment. The only difference is that in our simulation the melting speed is even faster, possibly because of the small supercell size used in the simulation and other complications in experiment including surface effects and a large pulse width used ($200$~fs).

\subsection{Validating  Nonthermal Characteristics in Laser Melting}

\begin{figure*}
	\includegraphics[width=\fitwidth]{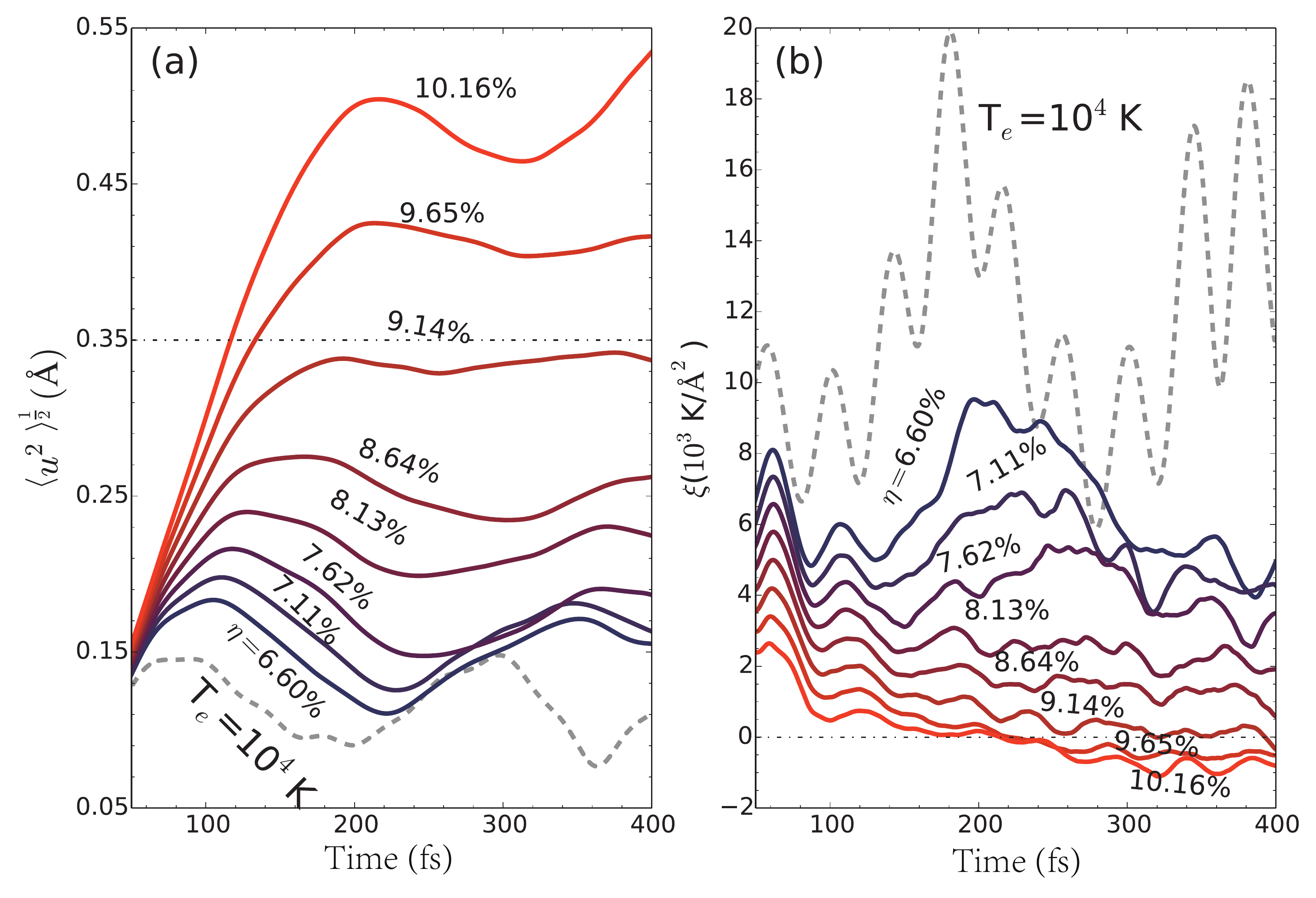}
	\caption{\label{VR}The (a) RMSD and (b) $\xi$ as a function of time under
		different laser intensities.}
\end{figure*}

We adopt a simplest model to evaluate the energy barrier for melting. We use a harmonic
potential to represent the interatomic interaction, similar to that in Ref.~\onlinecite{Lindenberg2005}. Since the total energy is conserved during melting, the increase in potential energy $\Delta E_p(t) \propto \left<u^2(t)\right>$ equals the decrease in ionic kinetic energy, $\Delta E_k(t) \propto \Delta T(t) = T(t_0) - T(t)$. Thus we obtain $\xi(t) \left<u^2(t)\right> \equiv \Delta T(t)$, where $\xi(t)$ is a constant. If melting occurs with $\left<u^2(t)\right>^{\frac{1}{2}} = R_c$, the temperature needed for melting is approximately $T_c \geq \Delta T = \xi {R_c}^2$, and the barrier is estimated as $k_B T_c = \xi k_B {R_c}^2$, where $k_B$ is the Boltzmann constant.

Figure \ref{VR} displays the evolution of RMSD and $\xi(t)$ under different laser intensities. We find that the maximum of RMSD increases as the laser intensity increases. A critical intensity $\eta_c = 9.14$\% is found when the RMSD just reaches $R_c = 0.35$~\AA. It agrees well with the threshold intensity of $5$--$10$\% for laser-induced melting found in experiment~\cite{Sundaram2002}. Consequently, as laser intensity increases, the nominal melting barrier $\xi(t)$ decreases [Fig.~\ref{VR}(b)]. For laser intensity $\eta = 10.16$\%, maximum $\xi(t)$ at $t\sim60$~fs is $2.4\times10^3$~K/\AA$^2$, meaning only a temperature $T = 294$~K is needed to reach melting with $R_c = 0.35$~\AA. Early works based on tight bonding models showed similar behaviors in laser-excited GaAs~\cite{Roland-prb02}. Thus Si melts at room temperature, which is much lower than the
thermal melting point of $1680$~K.

\begin{figure}
	\includegraphics[width=\fitwidth]{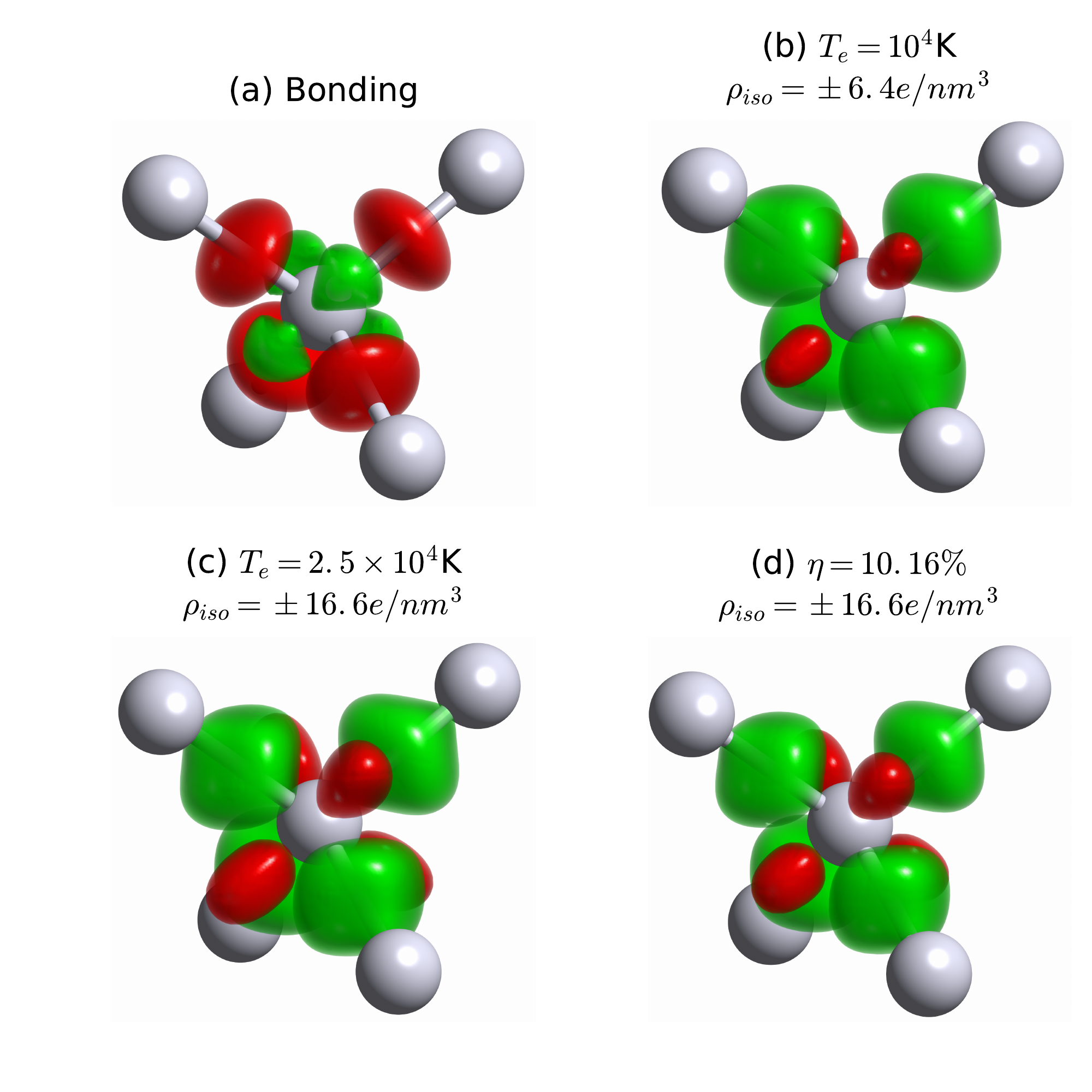}
	\caption{\label{DenChar} (a) Charge
		density difference between the ground state and the superposition
		of atomic charge densities. (b)--(d) Charge density difference
		induced by laser excitation in TTM-BOMD and TDDFT. The red (green)
		region represents the density increase (decrease) in charge density. }
\end{figure}

The shrinking melting barrier upon excitation is further attributed to laser-induced bond weakening. To illustrate its electronic origin, we display in Fig.~\ref{DenChar} laser-induced charge density difference $\Delta\rho$. Comparing $\Delta\rho$ for $\eta = 10.16$\% (Fig.~\ref{DenChar}(d)) with $\Delta\rho$ in the bonding state [Fig.~\ref{DenChar}(a)], we find that laser pulses induce electron transfer from the bonding state to antibonding state, thus significantly lowering the melting barrier. The amount of charge transfer can be evaluated by $\Delta n = (1/2N_a)\int_v|\Delta\rho| d^3r$, where $v$ is the volume of the unit cell and $N_a$ is the number of atoms in the the unit cell. As shown in Table~\ref{EC}, $\Delta n = 0.104$~e/atom is transferred from the bonding state to antibonding state with laser intensity $\eta = 10.16$\%, accounting for the barrier decrease in laser excited Si. Schultze \textit{et al}.~\cite{Schultze2014a} also reported that the absorption process produces a transfer of electron density from bonding to anti-bonding orbitals, though we use a different method based on the Fermi golden rule to build the initial excitation state.

\begin{figure}
	\includegraphics[width=\fitwidth]{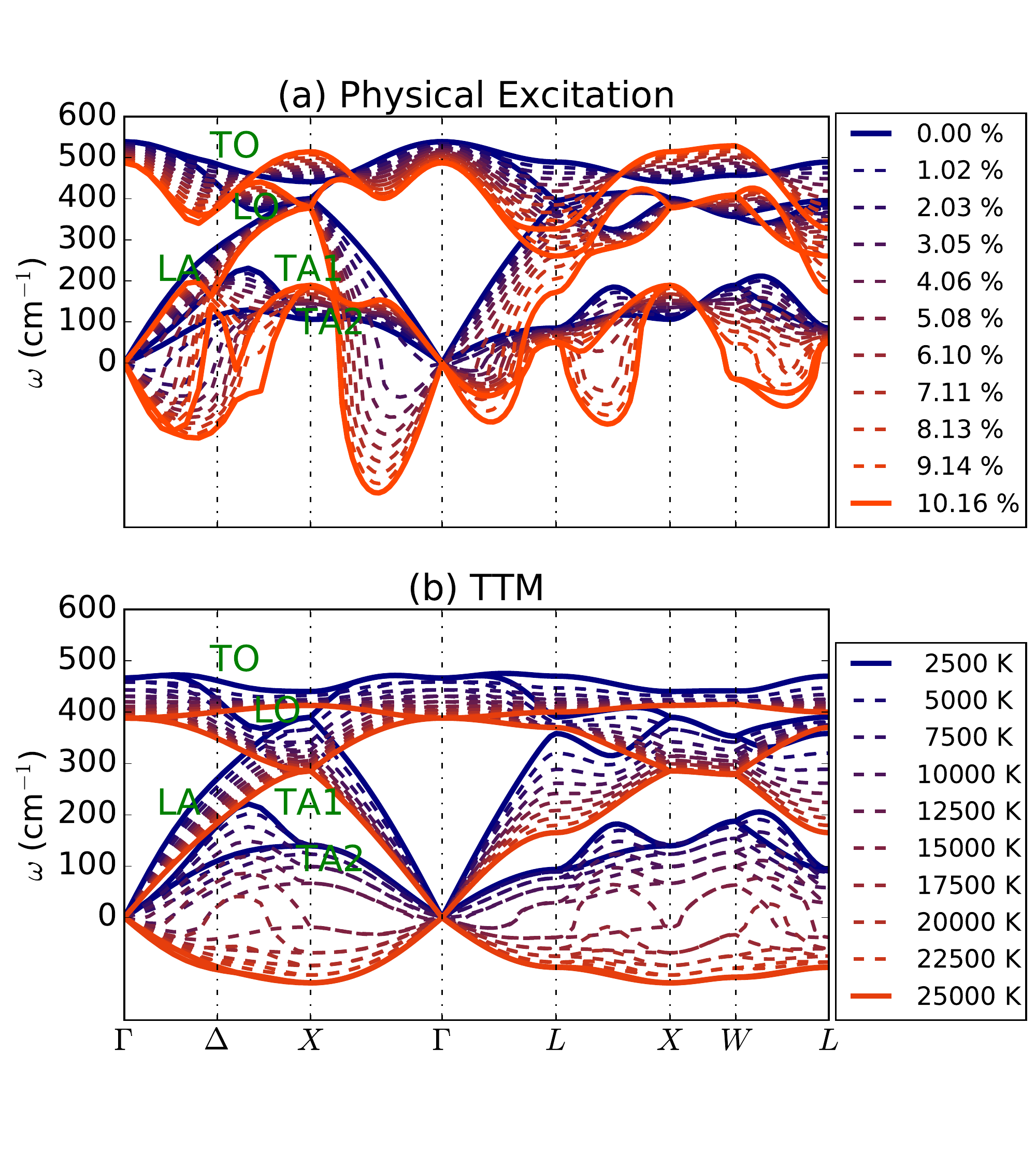}
	\caption{\label{Figure5}Phonon dispersion spectra of Si under
		(a) different laser intensities and (b) different electronic temperature.}
\end{figure}

Phonon dispersion analysis further confirms that upon laser excitation, both the longitudinal acoustic (LA) and transverse acoustic (TA) phonon modes destabilize [Fig. \ref{Figure5}(a)]. For laser intensity $\eta > 1$\%, imaginary frequencies show up all over the Brillouin zone, especially along the $\Gamma-X$ and $\Gamma-\Delta$ directions. Severe photon softening leads to ultrafast melting. Note that phonon instability calculated from physical excitation conditions is stronger than that in TTM models, where only TA modes are significantly softened [Fig. \ref{Figure5}(b)]. The latter is inconsistent with experimental analysis that all phonons are affected \cite{Lindenberg2005}. In TTM the LA mode is only slightly perturbed, while under physical excitations it is drastically destabilized, indicating Si during laser melting is different from normal liquid assumed in Ref. \onlinecite{VANVECHTEN1979422}, where the LA mode is largely maintained. Thus, the melting occurs not only through destabilizing the shearing mode, but also the stretching mode.

\begin{figure}
	\includegraphics[width=\fitwidth]{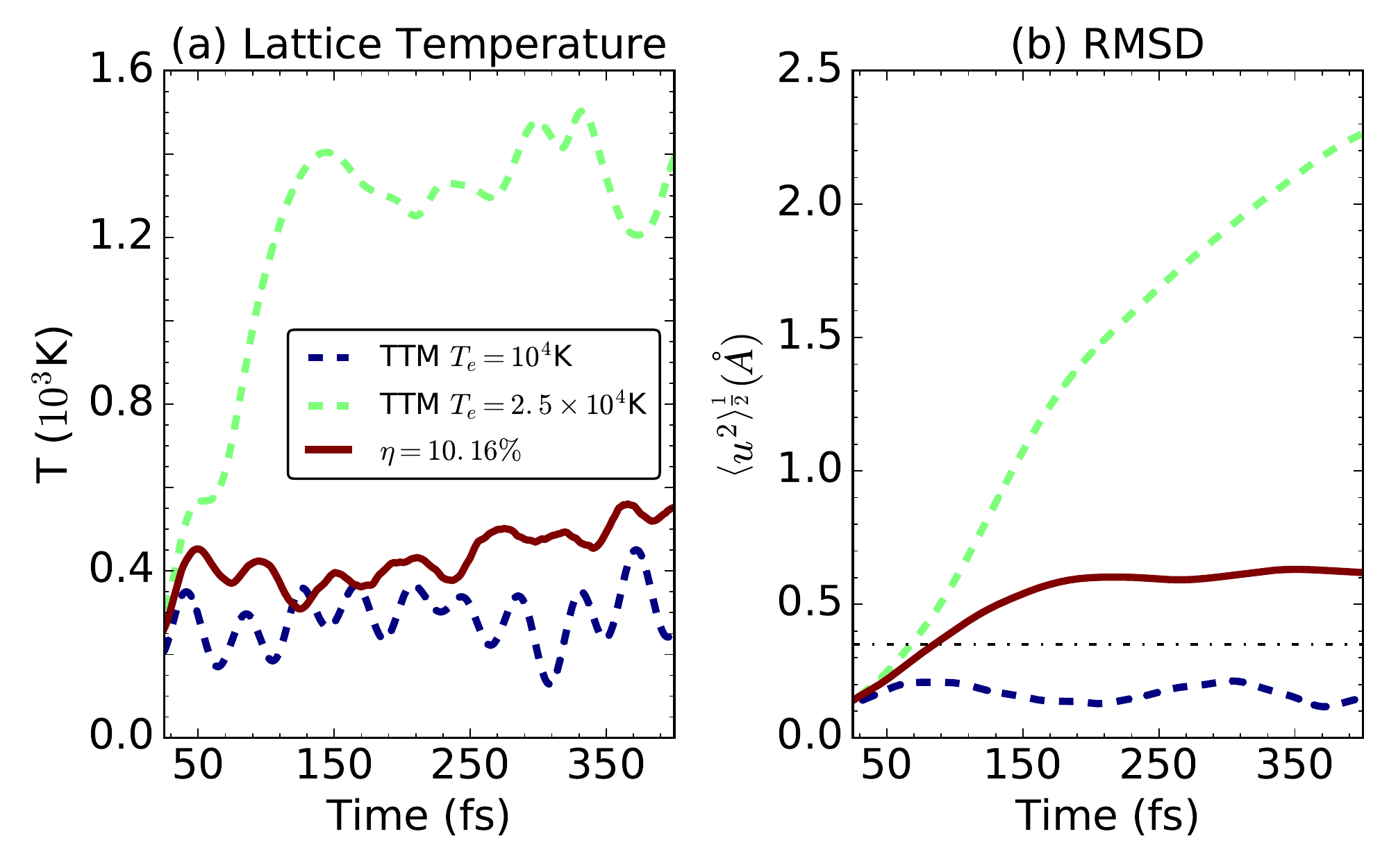}
	\caption{\label{BTBOMD}(a) Ionic temperature as a function of time. (b) The RMSD as a function of time.}
\end{figure}

We find that the widely used TTM based on the BO approximation yields radically different behaviors. The maximum laser energy in experiment is estimated to be $0.29$~eV per valence electron, similar to the energy gain with an electronic temperature $T_e = 10^4$~K (Table~\ref{EC}). However, no melting is observed with $T_e = 10^4$~K, whose temperature and RMSD are shown in Fig. \ref{BTBOMD}. The RMSD is only $0.1$~\AA, far below $R_c = 0.35$~\AA. This is resulted from insufficient excitation in TTM: the $\Delta n$ for $T_e=10^4$ K is only half that for $\eta=10.16$\%; see Fig.~\ref{DenChar} and
Table~\ref{EC}. Consequently, Si bonds are not sufficiently weakened and the melting barrier is overestimated. From Fig. \ref{VR}(b), the average $\xi_{T_e=10^4K}(t)$ is $1.3\times10^4$~K/\AA$^2$, much larger than $\xi_{\eta=10.16\%}(t)$. To reach $R_c = 0.35$~\AA, a temperature $T = 1600$~K is needed, close to the thermal melting temperature $1680$~K. Thus, there is no nonthermal melting in the TTM simulations.  

A previous study reported that Si melts with a much higher $T_e = 2.5\times 10^4$~K \cite{Silvestrelli1996}. In this case, the $\Delta n$ value is close to that for $\eta = 10.16$\%. However, the RMSD rapidly increases unreasonably to $2.3$~\AA\ in $400$~fs, much larger than $0.6$~\AA\ as observed in experiment~\cite{Harb2008}. More importantly, the lattice temperature rapidly increases to $T>1400$~K in $160$~fs, which is a common feature of TTM reported also for Ge melting \cite{Ji2013}, implying a thermal process. After all, the energy required to reach $T_e = 2.5\times 10^4$~K is $\sim4$ times larger than the energy input from laser pulses (Table~\ref{EC}), violating energy conservation law.

\begin{table}
	\caption{\label{EC} The total excitation energy and the charge
		transfer $\Delta n$. The experimental excitation energy is
		evaluated as $E = \eta_{exp} E_g$, where
		$\eta_{exp} = 11$\% is the experimental percentage of the
		excited electrons from Ref. \onlinecite{Harb2008} and $E_g = 2.6$~eV is
		the direct band gap of Si after scissor correction.}
	\begin{ruledtabular}
		\begin{tabular}{lcc}
			System & Energy(eV/electron) & $\Delta n$(e/atom)\\\hline
			This work ($\eta = 10.16$\%)       &$0.28$  &$0.104$ \\
			TTM ($T_e = 10^4$~K)               &$0.33$  &$0.045$ \\
			TTM ($T_e = 2.5\times 10^4$~K)     &$1.25$  &$0.112$ \\
			Experiment \cite{Harb2008}         &$0.29$  & - \\
		\end{tabular}
	\end{ruledtabular}
\end{table}

Thus, the melting is not caused by the ultrafast ionic temperature increase indicated by the TTM. The evidences of this thermal melting model originate from the drawbacks of TTM~\cite{SM}. Instead, the PA mechanism, i.e., plasma-induced bond weakening, seems to work well to explain the above results. The PA model assumes that laser energy is retained in the electronic subsystem,
and thus ultrafast melting is purely an electronic effect. 

\subsection{Energy Transfer Beyond the Plasmon Annealing Model}

However, we also show that the PA is also insufficient to fully understand ultrafast melting.  
In the melting process, the thermal velocity of ions $v_T = \sqrt{3k_BT/M}$ ($M$ is the atomic mass) is consumed to overcome the barrier. Thus the melting velocity predicted in PA, $v^{Melt}_{PA}$, would always be smaller than $v_T$. However, based on our \textit{ab initio} nonadiabatic simulations, the calculated ion velocity $v^{Melt}_{Cal}$ during melting is equal to or even larger than thermal velocity  $v_{T}$ at low temperature. For instance, the ionic melting velocity is calculated to be 1.8 \AA/ps at the lattice temperature $30$ K, which is much larger than the average thermal velocity of 1.4 \AA/ps. This phenomenon can not be explained by the PA model, and thus hints for a different mechanism, i.e., thermal accelerated plasma annealing (TAPA). In the TAPA picture [Fig.~\ref{NMPAModel}(c)], $v^{Melt}_{TAPA} = v^{Melt}_{PA}  + v_{el-ph}$, additional increase in melting velocity is induced by finite el-ph energy transfer $v_{el-ph}$. The $v^{Melt}_{TAPA}$ reproduces well the tendency of $v^{Melt}_{Cal}$, with an el-ph energy transfer $E_{el-ph}$ about $10$~meV/Si. Therefore we conclude that the small but finite energy transfer from electrons to the ionic degree of freedom is critical to fully and precisely understand ultrafast melting dynamics under laser illumination.

\section{Conclusion}
{ We have investigated laser melting of Si using real time TDDFT, where physical excitation conditions and nonadiabatic effects are naturally included. Without adjustable parameters the simulation shows nonadiabatic and nonthermal ultrafast behavior, in excellent agreement with experimental measurements in terms of laser threshold and decay of diffraction intensity. During melting the crystal lattice remains cold ($T<600$~K), suggesting a nonthermal phenomenon, which is further attributed to drastic laser-induced changes in bonding electron density and subsequent decrease in the melting barrier. Moreover, we show that the small but finite el-ph energy transfer, absent in the PA model, is key to induce accelerated melting dynamics at low temperature. Thus, we validate and extend the PA mechanism of nonthermal melting speculated 40 years ago, and provide additional insights about the ultrafast electron-phonon energy transfer. This approach could be extended to study other nonthermal phenomena of materials induced by laser illumination. }

\section{ACKNOWLEDGMENTS}
We acknowledge financial supports from MOST (Grants No. 2016YFA0300902 and No. 2015CB921001), NSFC (Grants No. 11222431 and No. 11222431), and CAS (XDB07030100). S.B.Z. was supported by the U.S. Department of Energy under Grant No. DE-SC0002623.

%
\end{document}


\title{Supplemental Materials for \\
Ab initio evidence for nonthermal dynamics in ultrafast laser melting
}
\author{Chao Lian}
\affiliation{Beijing National Laboratory for Condensed Matter Physics and Institute of Physics, Chinese Academy of Sciences, Beijing, 100190, P. R. China}

\author{S. B. Zhang}
\affiliation{Department of Physics, Applied Physics, and Astronomy, Rensselaer Polytechnic Institute, Troy, New York 12180, USA}

\author{Sheng Meng}
\affiliation{Beijing National Laboratory for Condensed Matter Physics and Institute of Physics, Chinese Academy of Sciences, Beijing, 100190, P. R. China}
\affiliation{Collaborative Innovation Center of Quantum Matter, Beijing, 100190, P. R. China}

\date{\today}
\maketitle


{We note that, the empirical two-temperature model (TTM) is successful in modeling the ultrafast laser-matter interaction, especially in metals where the electrons nearly follow the Fermi-Dirac distribution. Also, many improvements in TTM could consider the non-Fermi-Dirac distributed electrons  \cite{Wentzcovitch1992,Sun1994,Lisowski2004,Carpene2006}. However, if the TTM is combined with the Born-Oppenheimer (BO) approximation in the ab initio molecular dynamics (MD), $T_e$ is constant $\frac{\partial T_e }{\partial t} = 0$. The basic differential equation of the empirical TTM,}
\begin{equation}
C_e\frac{\partial T_e }{\partial t} = - C_l \frac{\partial T_l}{\partial t},
\end{equation} 
{requires that $\frac{\partial T_l}{\partial t} = 0$, where $T_l$ is the lattice temperature.
	However, in TTM-BOMD $T_l$ increases dramatically because of the free energy conservation. Thus, the TTM-BOMD yields a false el-ph energy transfer, which causes the thermal melting behavior in the TTM-BOMD.}

\begin{figure}
\includegraphics[width=\fullwidth]{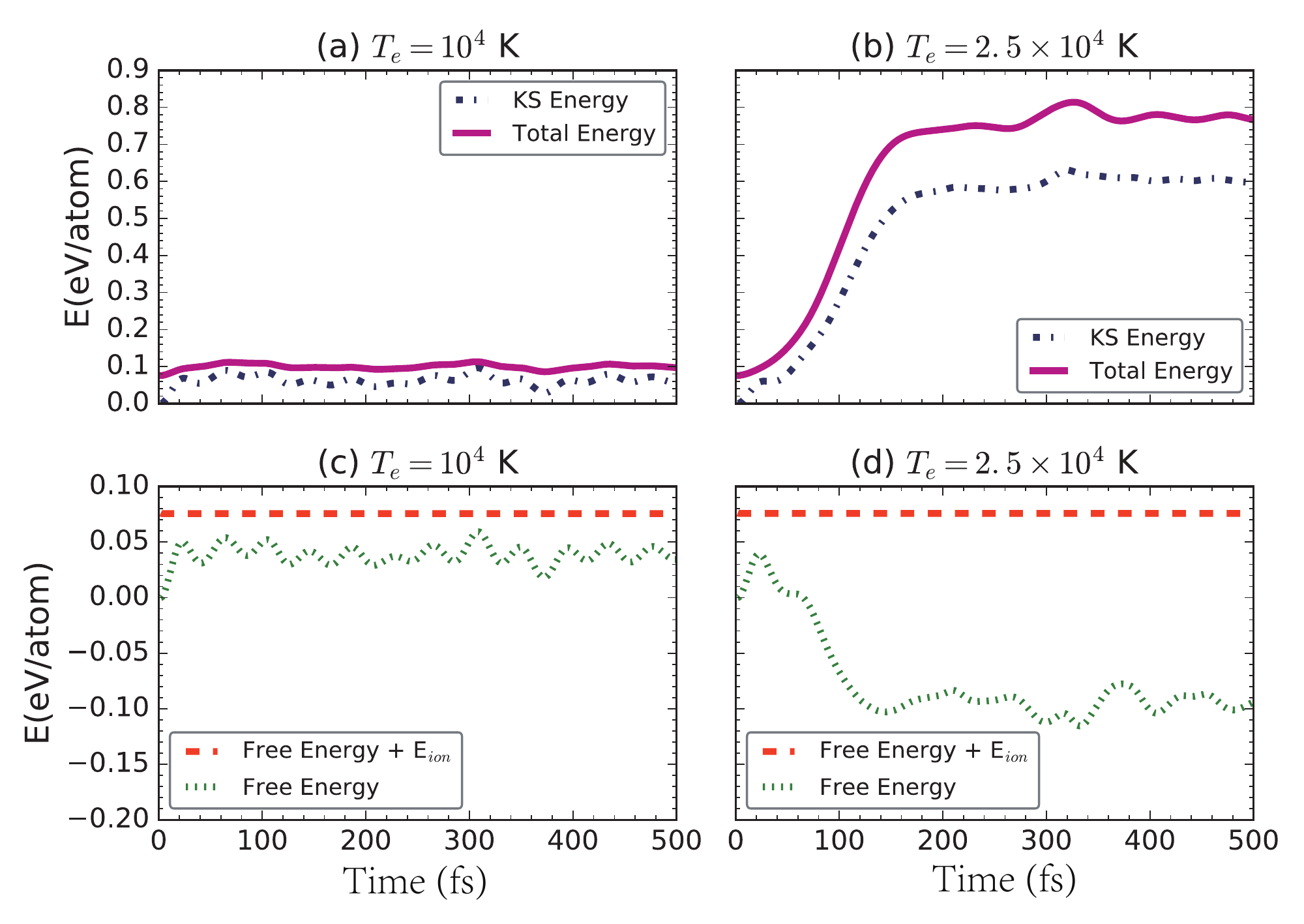}
\caption{\label{S1}(Color online) The total energy and Kohn-Sham energy as a
function of time in TTM simulations with (a) $T_e = 10^4$~K and
(b) $T_e = 2.5\times 10^4$~K, respectively. The KS energy at $t =
0$~fs is set to zero. The electronic free energy and total free
energy as a function of time in TTM with (c) $T_e = 10^4$~K and
(d) $T_e = 2.5\times 10^4$~K, respectively. The electronic free
energy at $t = 0$~fs is set to zero.}
\end{figure}

The failures of TTM in describing ultrafast laser melting root from the drawbacks of TTM. In the Born-Oppenheimer MD, the occupation number $f_i$ of each
state $i$ is fixed for $T_e \rightarrow 0$~K. Thus, the total
energy $E_{tot} = E_{ion} + E_{KS}$ is conserved, where $E_{ion}$
is the kinetic energy of ions, and $E_{KS}$ is the total Kohn-Sham
energy of electrons. It indicates a microcanonic ensemble.
However, for the cases $T_e \gg 0$~K in TTM simulations, $f_i$
changes violently with time (as the orbital energy varies with
respect to the Fermi level). Since the KS functional is not
variational with $f_i$, the $E_{tot}$ is not conserved, as shown
in Fig.~\ref{S1}(a-b). Instead, the forces are calculated as
\begin{equation}\label{TTMforce}
F = -\nabla \Omega,
\end{equation}
where $\Omega$ is the Mermin free energy $\Omega = U - T_e S$, and
$S$ is the entropy of electrons. Thus, $\Omega + E_{ion}$ is
conserved, shown in Fig.~\ref{S1}(c-d). Wentzcovitch et al.
\cite{Wentzcovitch1992} pointed out that it is a statistical
hypothesis originated from the BO approximation that $f_i$ can be described using $T_e$. However this approximation is not valid at very high
$T_e$, because the actual time evolution of $f_i$ should be
governed by time-dependent sch\"{o}dinger equation (here
time-dependent KS equation). Although the forces are calculated
reasonably using Eq. \ref{TTMforce}, the constraint condition of
the microcanonical ensemble is violated. The large total energy
shift originates from the presumed constant electron temperature.
According to real time evolution shows in Fig.~\ref{S1}(b), the
$f_i$ is far from the Fermi-Dirac distribution with a constant
$T_e$. The failure leads to nonphysical results of TTM as we
mentioned in the text.

%